# Plasma Generation by Household Microwave Oven for Surface Modification and Other Emerging Applications

Benjamin K Barnes,[1, 2] Habilou Ouro-Koura,[3, 4] Jesudara Omidokun,[3] Samuel Lebarty,[3] Nathan Bane,[3] Othman Suleiman,[3] Justin Derickson,[3] Eguono Omagamre,[5] Mahdi J Fotouhi,[5] Ayobami Ogunmolasuyi,[3, 6] Arturo Dominguez,[7] Larry Gonic,[8] and Kausik S Das[5, *]

[1]*University of Maryland Eastern Shore, 1, Backbone Road, Princess Anne, MD 21853 USA*

[2]*Department of Chemistry and Biochemistry, University of Maryland College Park, MD 20742 USA*

[3]*Department of Engineering, University of Maryland Eastern Shore, 1, Backbone Road, Princess Anne, MD 21853 USA*

[4]*Department of Mechanical Engineering, Rensselaer Polytechnic Institute, Troy, NY 12180 USA*

[5]*Department of Natural Sciences, University of Maryland Eastern Shore, 1, Backbone Road, Princess Anne, MD 21853 USA*

[6]*Department of Mechanical Engineering, Dartmouth College, Troy, NY 12180 USA*

[7]*Princeton Plasma Physics Laboratory, 100 Stellarator Rd, Princeton, NJ 08540 USA*

[8]*247 Missouri Street, San Francisco, CA, 94107*

(Dated: May 30, 2020)

# Abstract

A simple and inexpensive method to generate plasma using a kitchen microwave is described in this paper. The microwave-generated plasma is characterized by spectroscopic analysis and compared with the absorption spectra of a gas discharge tube. A Paschen-like curve is observed leading to a hypothesis of the microwave plasma generation mechanism in air. We have also demonstrated that this microwave-generated air plasma can be used in a multitude of applications such as: a) surface modification of a substrate to change its wettability; b) surface modification to change electrical/optical properties of a substrate; and c) enhancement of adhesive forces for improved bonding of polymeric microfluidic molds, such as bonding polydimethylsiloxane (PDMS) chips to glass covers. These simple techniques of plasma generation and subsequent surface treatment and modification applications may bring new opportunities leading to new innovations not only in advanced labs, but also in undergraduate and even high school research labs.

## I. INTRODUCTION

Plasma is ubiquitous in nature and readily observed as lightning on a stormy night, or as a static electric spark the moment before touching a door knob on a dry winter day[1]. It is also found naturally occurring in stars[2], solar winds[3], upper atmospheric lighting[4] and the awe generating lightning glow known as St. Elmo's fire[5] that sailors wondered about for centuries. Practical applications of plasma span a vast domain including, but not limited to, fusion energy generation[6–9], plasma TV displays[10], plasma enhanced chemical vapor deposition (PECVD)[11], sputtering[12], semiconductor device fabrication[13,14], substrate cleaning[15] and sterilization[16]. Plasma is also referred to as the fourth state of matter[17]. As the addition of energy to material in a solid or first state initiates a phase change to a liquid second state and then to a gaseous third state, adding a sufficient amount of energy to a gas may ionize the gas by ejecting electrons from the outer shells and/or through collisions of individual ions. Plasma, the fourth state[17], is thus composed of a conducting soup of energised charged particles, electrons and ionized atoms/molecules that are normally generated by the application of strong electromagnetic or heat energy. Once this highly charged gas is brought in contact with a surface, some energy is transferred from the plasma to the substrate surface, in turn changing the surface property of the substrate. Unfortunately, plasma instrumentation remains somewhat sophisticated and expensive, preventing its widespread use in teaching and research environments for institutions with small budgets. In the past some attempts were made to bring plasma research and analysis to high school and undergraduate teaching and research labs[18–20], however, they either involve gas discharge tubes that require high voltage and vacuum levels, or high temperature flames[21] for the proposed experiments, raising potential safety concerns and related issues in elementary labs. Beck et. al.[22] also computed trajectories of electrons in a weakly coupled plasma obtained from particle simulations. However, the goal of nearly all of these earlier theoretical or experimental papers for undergraduate students was to characterize plasma and experimental/numerical verification of the theory of plasmas. In a recent paper[23] popular science activity of forming plasma in a microwave with grapes showed that grapes act as spheres of water, which, due to their large index of refraction and small absorptivity, form leaky resonators at 2.4 GHz. When brought together, Mie resonances in isolated spheres coherently add up so that the aqueous dimer displays an intense hotspot at the point of contact. When this hotspot is sufficient to field-ionize available sodium and potassium ions, it ignite a plasma.

In contrary to the previous works, here we show that a simple plasma treatment apparatus can be constructed from an inexpensive household microwave oven and a modified vacuum flask and showed many modern application of this microwave-generated plasma. Although some previous works have shown that plasma can be sparked when a gas held at low pressure is subjected to microwave radiation[24–35], this technique has never received the attention it deserves when it comes to low cost laboratory technique for wide range of surface treatments. We demonstrate a number of cutting-edge applications of the plasma device that we have constructed which includes the modification opto-electrical properties of semiconductors, changing wettability of substrates, enhancement of bonding between certain polymers such as PDMS and glass and reduction of graphene oxide to graphene for electronics purposes etc.

Before going into the details of the microwave generated plasma generation process, we believe a comprehensive discussion about past works on microwave breakdown of air, similarities and differences between previously observed breakdown phenomena and our plasma generation observation is necessary to put our work in proper perspective. In general plasma generation process under DC conditions is relatively well understood where avalanche multiplication of electrons in a neutral gas are produced through electron impact ionization (or photoionization) when electrons or photons with sufcient energy collide with neutral atoms and molecules present in the medium[36]. However, under the influence of an oscillating electric field instead of a DC field the ions and electrons would oscillate back and forth in between the electrodes and for a very high frequency only a very small number of charged particles would be able to reach the boundary electrodes reducing the diffusive loss of electron energy.

The first serious investigation of microwave breakdown in air was conducted by Herlin and Brown[32]. Theoretical basis of their analysis consists of solving the continuity equation for the electron number density in their custom made cavity geometry, accounting for the source of electrons as impact ionization and loss terms by diffusion and attachment corresponding to the gas in use. Their custom made microwave source was a continuous wave tunable magnetron that operated at about 3 GHz. In order to isolate the microscopic phenomena leading to breakdown they used specifically designed cavities (i.e., cavities that resonated at the $TM_{010}$ mode) so that gas pressure and electric field strength were at all times well-known. A radioactive source was placed near the discharge that provided a small amount of ionization in the cavity to start with, and the microwave field in the cavity was increased until the cavity suddenly began to glow (at which time the microwave field abruptly decreased). The radioactive source in their case was the supplier of the triggering electrons. They found that the breakdown field was smaller at about 3 Torr than it was at higher or lower pressures under these conditions. This minimum also appeared to be a weak function of cavity size. According to the results they have also found that the dominant mechanism for breakdown of a low pressure gas at microwave frequencies is ionization by collision of electrons with neutral gas molecules rather than loss by diffusion to the walls of the discharge tube.

Afterwards considerable research was done using fundamentally the same apparatus. Gould and Roberts[33] also used a cavity resonated at the $TM_{010}$ mode and a 85 millicurrie, cobalt 60 radioactive source of was placed on the cavity in order to ensure the presence of sufficient number of electrons in the discharge region even before the microwave was turned on. They also concluded that under these conditions the mechanisms controlling breakdown in air were electron generation by impact ionization, and loss by attachment and diffusion. Good agreement between the theory and their experiment over a wide range of parameters verified the assumptions and the values of the coefficients used in the theory. Of note is the extension of the theory to include as a loss mechanism for electrons the attachment of electrons to neutral molecules, thus creating negatively-charged molecules (of note also is that recombination, the coalescence of an electron with an ion to form an atom, is negligible).

MacDonald, Gaskell, and Gitterman[34] extended the range of frequencies studied, and investigated air, nitrogen, and oxygen independently. With the exception of some X-band cavities, all of their cavities were also cylindrical and resonant in the $TM_{010}$ mode. They also reported the necessity of having a 5 millicurie cobalt 60 gamma-ray source next to the cavities to produce sufficient ionization to get repeatable results for pulsed waves and 5 microcurie source for continuous wave process. Their measurements agreed with previous measurements. They critiqued the previous theory by noting it assumed an electron-molecule collision frequency that was independent of electron energy, when it had long been known that the electron-nitrogen collision frequency was in fact strongly dependent on electron energy. While unable to present an improved analytic theory, they were able to present an alternative representation that allowed an empirical method of calculating breakdown fields in a very large variety of conditions.

MacDonald, Gaskell, and Gittermans work was extended to form the final chapter (and appendix) of MacDonalds book[30]. As had been the case in Herlin and Browns work, in order to achieve reproducible results a radioactive source (5 millicourie $Co^{60}$) was placed again near the cavity in this work. Bandel and MacDonald[35] further measured breakdown electric fields in air, $H_2O$, and air plus $H_2O$ at 3.06 GHz. In their case triggering electrons for the discharge were supplied not by a radioactive source, but by photo-electric emission by ultraviolet light, introduced into the cavity from a synchronized spark discharge in room air. Care was taken to attenuate the uv as needed to avoid lowering of the breakdown threshold. They found that even in the presence of triggering electrons from photo-electric emissions the breakdown field strengths for mixture of air plus 17.2 Torr of water vapor were up to about 25% greater than for dry air at the same total pressure, thus quantifying the effect of water vapor molecules in microwave plasma generation in their resonant cavity geometry.

In case of breakdown fields required when using very short microwave pulses MacDonald[30] also confirmed that in the absence of an electron from some auxiliary source, the breakdown field measured would be very large, since the initial electrons would have to be provided by field emission or other surface effects.

In the context of all these previous serious works on microwave breakdown of air[30–35], we would like to point out that there are four distinct differences between the previous studies and our present work: a) minimum pressure required for plasma initiation in our chambers is orders of magnitude less than that of the previous ones, i.e, in the order of $\sim 10^{-2}$ Torr in our experiments in comparison to the order of $\sim 1$ Torr in the previous studies, b) minimum electric field required in our studies is $\sim 10^1$ V/cm, which is the standard average electric field present in a kitchen microwave, in comparison to $\sim 10^2$ V/cm or higher in the previous studies mentioned. Moreover, we used fixed power, fixed frequency microwave oven and we didn't vary average electric field as done previously, c) our plasma chamber is not specifically designed as a resonating cavity for any specific mode as done previously, rather, two different arbitrary shaped glass flasks and two different microwave ovens of different sizes produced similar qualitative results; and lastly, d) no external source of triggering electrons either in the form of radioactive sources or synchronized photo-electric emissions were necessary to generate plasma in a microwave oven as seen in the video[37] even in much lower electric field environment.

## II. PLASMA GENERATION BY A KITCHEN MICROWAVE

### A. Materials and methods

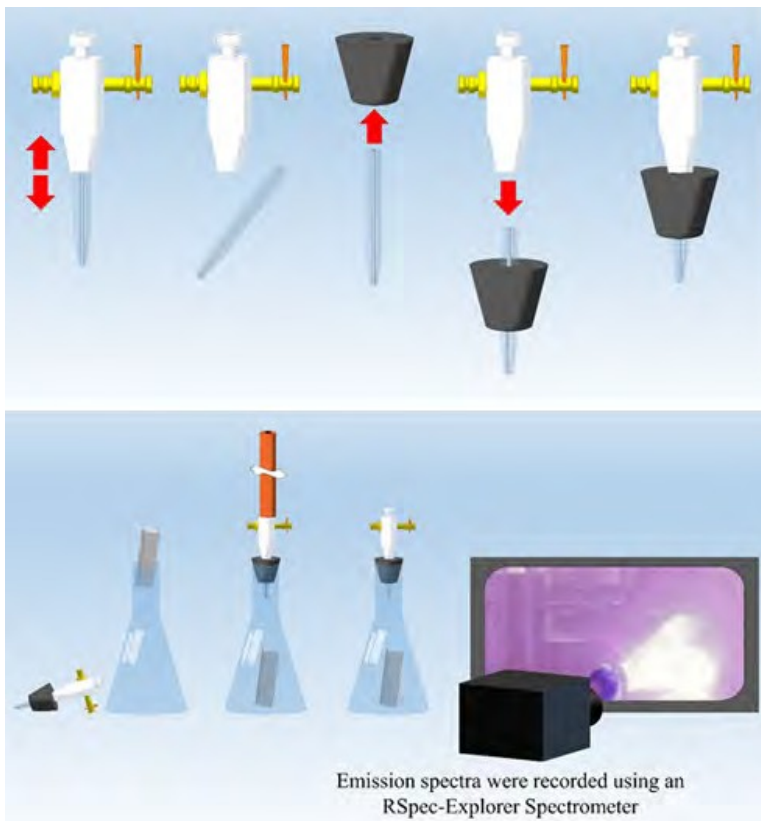

FIG. 1. A schematic of the construction of the sample holder for the microwave-generated plasma is shown in this figure. One of the crucial parts needed is a simple air-tight valve. An inexpensive one can be fabricated from specialty parts as shown here, by modifying a PTFE and glass burette tip and rubber stopper. Bottom left to right: the sample is inserted into the Erlenmeyer flask; the vacuum valve is opened and the vacuum pump is used to evacuate the flask; the valve is closed and the vacuum hose removed; the flask is transferred to the microwave oven, and as the microwave is turned on, plasma is sparked to treat the surface for the desired length of time[37]. To characterize the spectrum of the plasma, this process is repeated without a sample while placing a UV-Vis spectrometer in front of the microwave window.

The main component of this low-cost plasma etching system is a regular kitchen microwave oven and a vacuum flask/sample holder. Any robust, vacuum-gauge glass container will work, but we have found the Erlenmeyer flask is the simplest and least expensive. The flat bottom of this glassware gives extra stability during sample preparation which may be useful in a teaching environment. A valve must be attached to the flask stopper which can be opened while evacuating the chamber and closed again to seal and hold the vacuum. Speciality microwave-resistant valve parts can be purchased, but here we have sought a frugal alternative first. The ideal substitute for this part was determined to be the detachable two-part tips found on many burettes (Fig.1). These parts consist of an all PTFE housing and stopcock into which a glass tip is loosely inserted. The glass tip of the burette can be easily removed by twisting it slightly. Next, a rubber stopper with a hole in the center, such as those for inserting thermometers into reaction vessels, is used. This part is selected to fit the mouth of the Erlenmeyer flask used above. The glass burette tip above is then inserted through the bottom of the stopper until it protrudes from the top by about 10mm. The PTFE stopcock housing is then placed back over the protruding burette tip. All of the edges of this system are then sealed with regular silicone glue. The seal is sufficient to hold a vacuum for many hours. We have also found that a properly matched valve connector stopcock stopper for vacuum desiccators can serve this purpose also (Fig.2). The main point here is that the design of the container is not limited to the ones described here, rather using any microwavable vacuum container with a stopcock valve should produce plasma. A vacuum hose can be directly attached to the top of the valve allowing the system to be evacuated to various pressures with a small vacuum pump. Once the properly evacuated flask is put in the microwave and the oven is turned on, after a few seconds glowing plasma forms inside the flask (please watch the video of plasma generation[37]). The time difference between the first observation of plasma and the time of turning on the microwave oven, i.e., the plasma initiation time, is recorded for different vacuum pressures. Furthermore, by flushing the evacuated system with various gases like argon, nitrogen, oxygen etc. this system can be used to determine the effect of different plasma chemistries. A schematic of the whole process is shown in Fig.1. In all of the following experiments, the substrate or sample is placed in an Erlenmeyer flask, the vacuum hose is attached, and the pressure is lowered with a vacuum pump to the desired value. The valve is then sealed and the flask transferred to the microwave oven. Once the microwave oven is turned on after a short delay the plasma sparks, and then the plasma is allowed to glow for the desired amount of time.

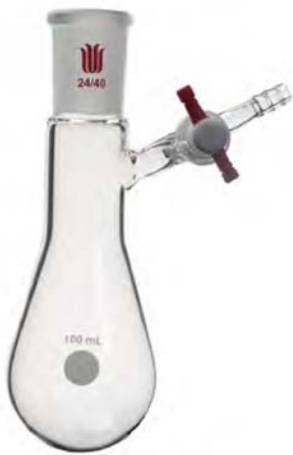

FIG. 2. An alternative microwavable vacuum container to generate microwave plasma.

### B. Characterization of the microwave-generated plasma

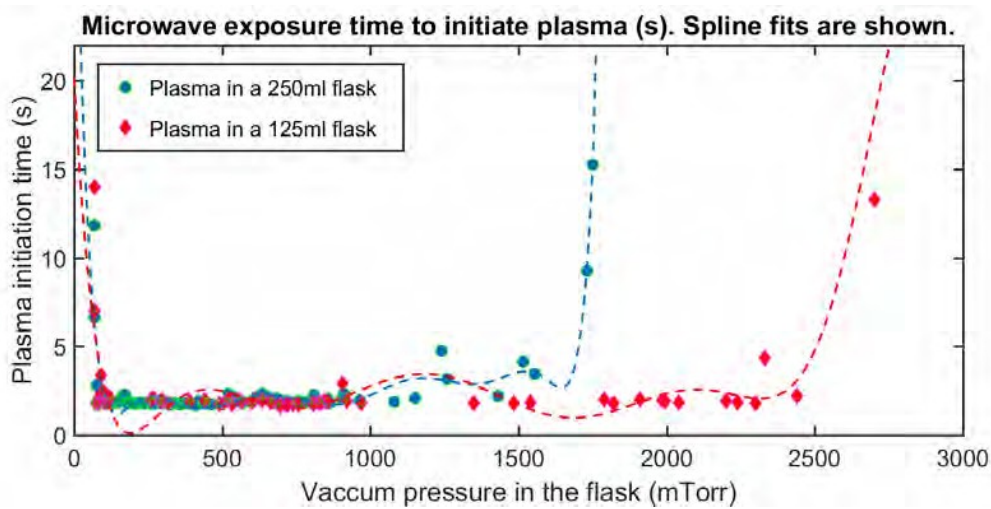

FIG. 3. Plasma initiation time measured and plotted as a function of vacuum pressure in two different flasks. A Paschen-like curve is observed when the pressure of the flasks are reduced and placed in the microwave oven to record plasma initiation time. At ambient pressures and high vacuum, the initiation time grows asymptotically, while in the intermediate regime, plasma sparks consistently within a few seconds. If the pressure in the flasks becomes too low, higher initiation time is observed pointing towards low number density of molecules in the flasks and corresponding lower collision cross section. Data for two flasks, one 125 ml and the other 250 ml show similar asymptotic behavior.

The plasma generated by this method is analyzed using two simple but revealing techniques: a) recording plasma initiation time as a function of the vacuum pressure, which is also a measure of particle density at constant temperature, and b) measurement of UV-visible emission spectra of the plasma glow. A Paschen-like graph[38,39] is generated by plotting the plasma initiation time with the vacuum pressure of the flask to varying level (Fig.3). A trough-shaped curve resulted from this analysis, showing asymptotic tendencies at high and low vacuum pressures and a sweet-spot regime in between in which the plasma sparks within a matter of seconds. To generate a classical Paschen curve in a gas discharged tube the breakdown voltage (or energy required for plasma initiation) between two electrodes is normally recorded as a function of pressure. However, in our system, energy required for plasma generation is supplied through constant power kitchen microwave, and the time the flask is exposed to microwave radiation determines the amount of energy supplied to the system. So, in our case we have plotted the microwave exposure time to initiate plasma as a function of the pressure of the flasks.

Although Herlin[32] or MacDonald's[30] theory based on the dynamics of impact ionization and diffusion loss were able to produce good match between their theory and experiments done in their chosen custom made resonant cavity geometry in presence of triggering electrons, our experimental results in a very different experimental conditions as described in the introduction section raises many unexplained questions that require fresh and in depth investigation. One such important question is about the origin of primary electrons in our set up in absence of any radio active source or photoemitters, since without a source of triggering electrons in a low electric field environment it is nearly impossible to start the electron avalanche process. Although actual source of triggering electrons is still unknown and further studies are needed, we hypothesize that the presence of stray water vapor molecules along with microwave field distortion near some edges, or cosmic ray dependent ionization may have some effect on the generation of primary electrons in a flask under microwave radiation. We believe that this is a plausible hypothesis since Bandel and MacDonald[35] also reported that even in the presence of a radio active source as a supplier of sufficient triggering electrons, presence of water vapor may affect the plasma generation process significantly (~ 25%). Water is a polar molecule[40,41] with dipole moment $p$ ( $|\vec{p}| \sim 6.2 \times 10^{-30}$ C.m) and it is well known that an instantaneous electric field can generate a torque to align these polar molecules in the direction of the instantaneous field (Supplementary Fig.16,17). In the oscillating electric field of a microwave water molecules are thus supposed to wobble/spin in synchronization with the field, in turn creating highly energized oscillating/rotating water molecules as shown in Fig.4. An order of magnitude calculation of the torque $\vec{\tau} = \vec{p} \times \vec{E}$ produced by a household microwave with frequency[42] $f = 2.45$ GHz and amplitude of electric field $|\vec{E}| \sim 2 \times 10^3$ V/m, shows that the microwave electric field may generate an enormous angular acceleration $\alpha$ of a water molecule ($|\vec{\alpha}| = \vec{\tau}/I \sim 10^{20}$ rad/s$^2$), $I \sim 3 \times 10^{-47}$ Kg.m$^2$ being the moment of inertia[43] of water molecule about the specific rotation axis. In actual scenario, however, the dynamics of asymmetric top type rotating water molecules are much more complex and may give rise to additional rotational instabilities, such as classical spin flipping, as observed in our zero gravity parabolic flight experiment done with a toy water molecule[44]. Moreover, it is also reported that intermolecular bonds get stretched in rapidly rotating molecules and there can be additional effects of electron screening to weaken the bond[45]. The effect of time varying magnetic fields on the diamagnetic water molecules[46], presence of cosmic rays, other sources of emissions etc. may also bring additional complexity to the dynamics. Whether these highly energized water molecules colliding with other atoms/molecules or other sources are capable of producing primary electrons is subject to debate. We think that it is still an open question and we invite other researchers to explore these unanswered questions.

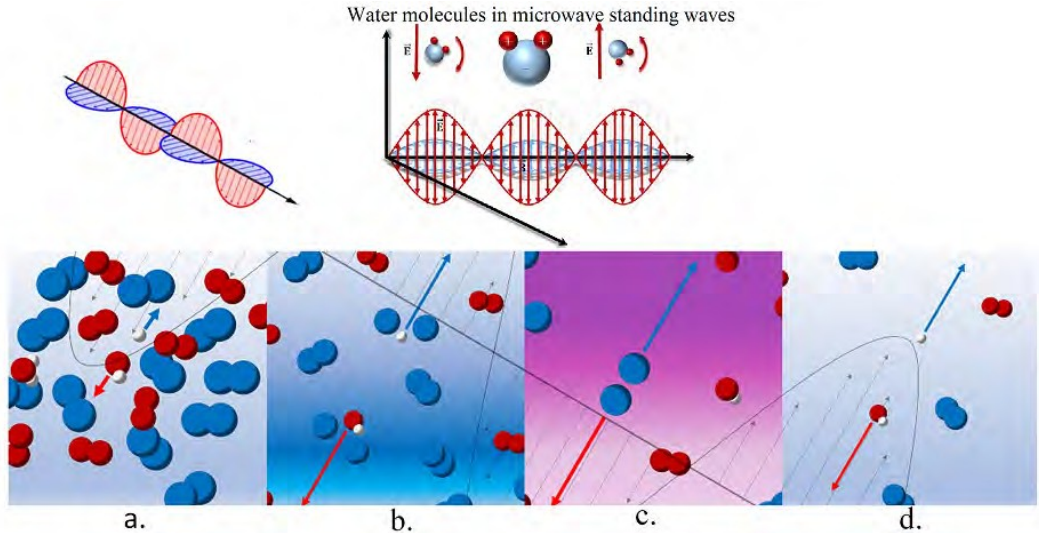

FIG. 4. The figure shows a possible mechanism of kitchen microwave generated plasma formation. Firstly, it is well known that electromagnetic waves in a standard kitchen microwave form a standing waves with a wavelength of 12 cm in a 2450 MHz. The electrical field component of the imposed electromagnetic wave excites the polar molecules like water. Positive charge centers of the polar molecules would experience a force/torque in the direction of the electric field while the negative charge centers would experience that in the opposite direction. Sufficiently strong and high frequency electrical field component of the applied microwave would be able to spin/vibrate the water molecules very fast and may eventually tear the negative and positive charge centers apart, most likely while hitting an atom/molecule/cosmic ray, and creating primary electrons and positive ions in the flask. Secondly (a-d), these primary electrons would accelerate opposite to the instantaneous applied electric field and collide with other molecules present in the flask on their way forward. If the electrons gather enough momentum before the collisions they will be able to knock out secondary electrons from the molecules/atoms they hit; in turn ionizing them also. These secondary electrons would then start to accelerate and may generate more and more electrons along the way creating a chain reaction. This is the pathway of the Townsend-type avalanche[38] we normally observe in a gas discharge plasma. A comparison of the plasma generated in a gas discharge tube and that in a microwave shows similar absorption spectra and supports this mechanism of microwave plasma generation.

Once the primary electrons are generated, it is plausible to assume that they will accelerate in the opposite direction of the instantaneous electric field and collide with any atom/molecule that come in their way creating secondary electrons and ions. Based on their charges these electrons and ions accelerate either towards, or in the opposite direction of the electrical field. A simple calculation shows that in response to the electric field the rate of kinetic energy gain in electrons is much higher than that of the corresponding ions. In case an electron of mass $m_e$ and charge $|q_e|$ is generated by collision along with a massive ion of mass $M_i$ and charge $|q_e|$ and both of them accelerate in an average electric field $E$, the ratio of kinetic energies of an electron and the ion can be obtained as $K.E_e/K.E_{ion} = M_i/m_e$. As $M_i/m_e \gg 1$, it is clear that electrons would be much more energetic in comparison to the massive ions in ionizing the gas in the flask and starting the cascading chain reaction leading to plasma. It is also well known[30] that the ion-molecule collision frequency is about $\sqrt{(m_e/M_i)} \ll 1$, so one can expect that the rate of ionization due to electron-molecule collision will be orders of magnitude higher than the rate of ionization due to ion-molecule collisions. It is well known that in electron-molecule collision frequency varies linearly with both the speed of the electrons and the electron-molecule collision cross section[30,47]. So, we hypothesize that speed of the electrons and electron-molecule collision cross section are two important parameters in the plasma generation process in a kitchen microwave. Since pressure in the plasma chamber at ambient temperature is directly proportional to the number density of the constituent particles, one can control the mean free path by lowering the number of constituent gas molecules or by lowering the pressure. At ambient pressure the number of constituent gas molecules in the flask is quite large and hence mean free path is too short (as mean free path is inversely proportional to the gas density). So, at higher pressure it would be difficult for the electrons to accelerate for a long distance and gain enough momentum (and kinetic energy) in between two successive collisions and cause an impact ionization chain reaction. Therefore plasma won't spark in this regime. As the pressure is decreased though, number density of particles decrease and the mean free path becomes sufficiently large for the knocked out electrons to gain enough momentum (and kinetic energy) to cause the nearby atoms to ionize upon collision, depending on their ionization energies. This is a possible mechanism that creates a sky-blue plasma which is visible in the first 1-2 seconds after ignition (please see the supplementary figures for illustrations). Further complexity in the collision mechanism can happen due to the collision of electrons and vigorously rotating water vapor molecules also. However, the chain reaction continues and molecules with higher ionization energy becoming ionized next due to collisions with electrons previously created. This regime is characterized by a deep purple and then pink plasma ball. When the vacuum pressure is decreased further to 68 mTorr or less for a 250 ml flask and 69 mTorr or less for a 125 ml flask, the collision cross section of electrons with molecules becomes too small for plasma to spark, and hence the plasma ignition time tends towards infinity in this asymptotic limit. Thus, we believe that two competing mechanisms are at interplay here: a) kinetic energy of the electrons accelerating in an electric field to knock out further electrons from atoms/molecules and b) collision cross section of the electrons and gas molecules in the flask. These two dominant regimes are visible in two asymptotic limits in Fig.3 as explained above. In the sweet spot the collision cross section and the kinetic energy of the electrons are just right to initiate plasma. We have two different kitchen microwaves and the same overall characteristic remains the same, although the plasma initiation times may vary slightly depending on the size and geometry of the microwave ovens.

UV-Visible absorption spectra of plasma are also recorded using a desktop CCD spectrometer (Exempler, B&WTek) and is shown in Fig.5. The spectra of the microwave plasma clearly show the dominant nitrogen and oxygen peaks indicating ionization of these molecules. To test our hypothesis of the microwave plasma generation we also built a gas discharge tube with fixed electrodes, generated air plasma in it and compared the spectra obtained from the gas discharge tube with that generated by the microwave. The reason behind this comparison is that gas discharge plasma is very well studied and the plasma generation mechanism of a gas under a constant strong electric field leading to Townsend avalanche is well known[38,48]. The striking similarity of both the spectra (Fig.5) lead us to confirm that although the source to energise gas is different in two cases: a) electric field in a gas discharge tube, while b) electromagnetic radiation in microwave-generated plasma, the chain reaction is somewhat similar as described in Fig.4.

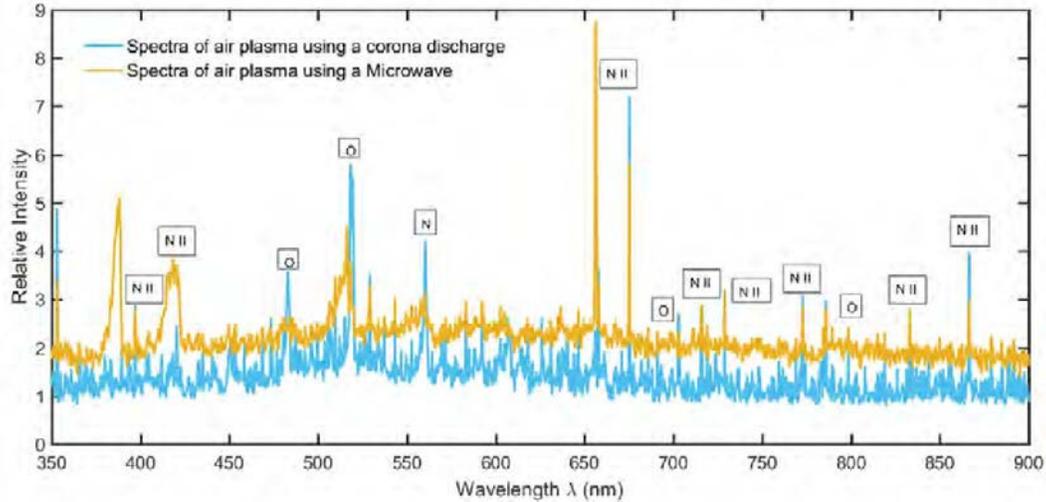

FIG. 5. The microwave-generated plasma spectra is compared with the corona discharge (gas discharge) tube air plasma spectra. Nitrogen and oxygen lines are visibly identified in both the spectra.

## III. APPLICATIONS OF MICROWAVE PLASMA

Plasma treatment is a well-known technique to remove any contaminant, change surface energy, modify electrical and thermodynamic properties of a substrate for various purposes[49]. A change in the surface energy of a substrate can easily be ascertained by measuring the contact angle of a liquid on the surface. The contact angle of a liquid drop, i.e., the angle measured where a three-component interface exists between a liquid drop, a solid surface, and the ambient air is measured by first taking cross sectional pictures of the drop using a camera and then analysing the contour of the drop using a software. Normally the designation of a hydrophobic material is reserved for surfaces with a liquid contact angle of greater than 90°s. Contact angles of less than 90° indicate a hydrophilic material with good wetting properties.

### A. PDMS surface modification by plasma treatment

A flat PDMS substrate is prepared by mixing the PDMS liquid elastomer and the curing agent (Sylgard 184) in 10:1 ratio followed by degassing and incubating for a couple of hours resulting in a consistent, bubble-free soft solid block. Soft cured PDMS blocks are then cut to shape and placed in the vacuum flask. The flask is evacuated to (500 mTorr) and placed in the microwave for plasma treatment. Once the plasma sparks, the PDMS blocks are treated for 2, 3, 4, and 5 seconds respectively. The relative change in surface energy is evaluated by measuring the contact angle of a water micro-droplet placed first on an untreated and then on the treated surfaces using a micropipette. We have used the open source image processing software ImageJ with DropSnake plugin to measure the contact angle of the sessile micro-droplet[50]. Fig.6 clearly shows that the contact angle of water on the microwave-plasma treated PDMS sample increases nearly linearly with the treatment time indicating a direct relationship between the change in surface energy and microwave-plasma etching. Surprisingly this observation is exactly opposite to the effect observed in regular gas discharge plasma treated PDMS surface where in absence of microwave just oxygen plasma lowers the contact angle and makes the substrates more wetting[51]. To further investigate this opposite effect of microwave plasma and to understand what role the microwave plays in this reversal of contact angle we have investigated the effect of microwave on a PDMS substrate with and without plasma Fig.6. It shows that while microwave (without plasma) has some small effect on making the PDMS surface less wetting as the contact angle rises from 63° to 70° with 10 s treatment time, the effect of microwave plasma is much larger as the contact angle changes from 63° to 159° with just 5 s of plasma treatment. We propose that the adsorbed water molecules on the surface of the PDMS substrate plays a crucial role here. One plausible explanation is that the surface adsorbed thin film of water gets heated up by the microwave component and simultaneously a rapid interaction with high density ions through the plasma contacts embeds ions on the soft-baked PDMS substrate surface leading the substrate towards superhydrophobicity. This microwave-plasma treatment technique is thus useful for making hydrophobic surfaces for water-resistant coatings and other applications, that can't be achieved through gas discharge plasma treatment, as well as for bonding PDMS to glass substrates and other applications, as investigated in detail in a later section.

**Surface modification of ZnO thin films to change its opto-electrical property.**

Plasma treatment of a metal oxide thin film can also change its opto-electrical property. A film of ZnO particles was prepared as previously described[52]. Briefly, a suspension of ZnO particles were deposited on a glass microscope slide from a 60% (v/v) solution of ethanol. The film was air dried for several minutes and then annealed on a hot plate by ramping the temperature from ambient to 500 °C in 7 minutes. Previous work has shown that as the ZnO film cooled a layer of ambient moisture and gases, as well as ethanol which is trapped in the bulk microstructure of the ZnO, deposits onto the ZnO surface[53]. This is verified by a strong OH band centered at 3400 cm$^{-1}$ in the FTIR spectrum of the film. The sample and substrate were transferred to the vacuum flask. The flask was then evacuated and the whole assembly was placed in the microwave oven as before. The plasma was sparked and allowed to interact with the ZnO surface for 4, 8, and 15 seconds. FTIR analysis of the treated surfaces in Fig.7 show a decrease in the overall intensity especially of the OH band (centered at 3400 cm$^{-1}$). Our previous work[52] has shown that such modification of polycrystal and nanocrystal semiconducting films plays an important role in the electronic properties of the material.

**Reduction of graphene oxide to graphene**

Graphene is a 2D wonder material[54–56] with many potential applications ranging from flexible electronics, energy storage, sensors to molecular sieving[57–60]. One hurdle in further developing graphene applications commercially however is the difficulty in preparation of bulk graphene in mass scale. One way of producing graphene is chemical vapor deposition (CVD), which is expensive and requires expensive instrumentation. Graphene is most easily produced in bulk by the reduction of graphite oxide solutions, but this can be time-consuming

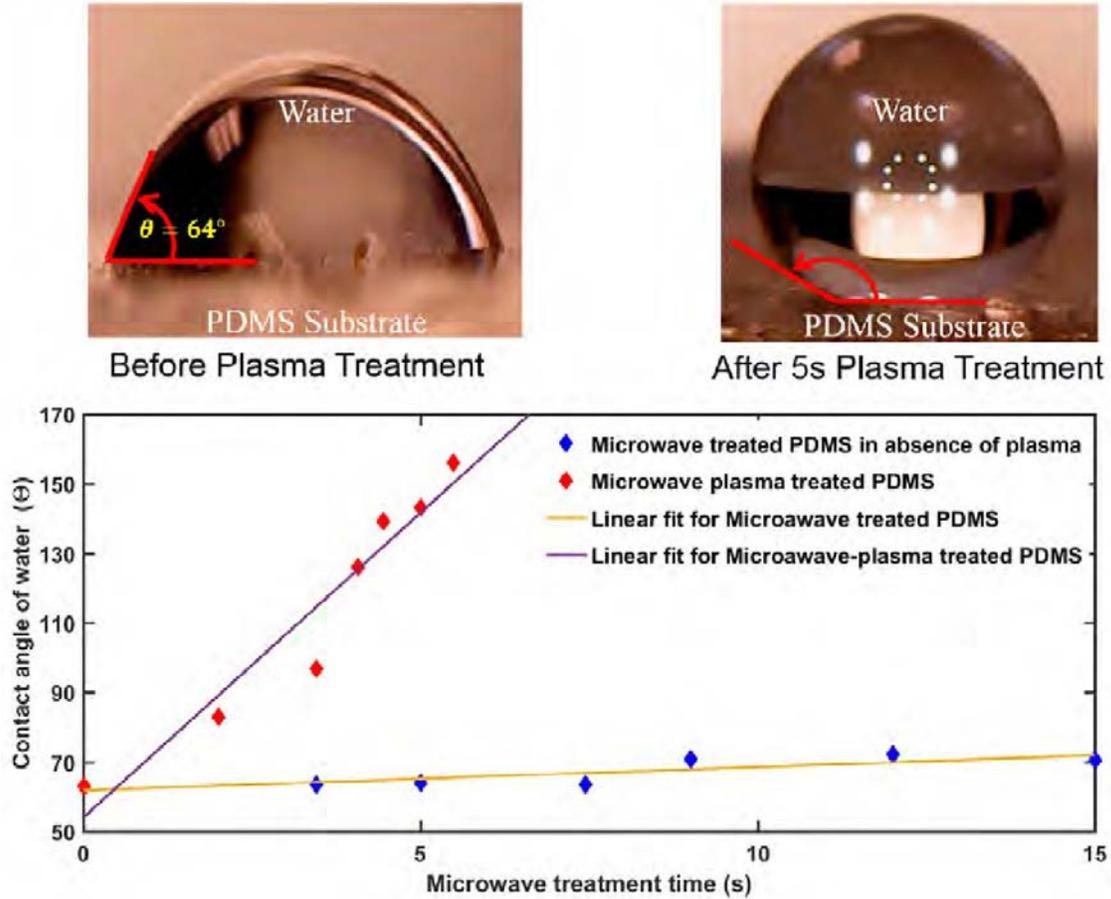

FIG. 6. Change in contact angle of water in a microwave oven, with and without plasma. Droplet pictures at the top show that increasing contact angle with extended plasma treatment time by adding sufficient surface energy of the soft cured PDMS polymer substrate. To differentiate the effects of microwave and microwave plus plasma one can choose a suitable pressure (say 2000 mTorr) and put the substrate in a 250 ml flask. As this pressure is above the higher asymptotic limit of the Paschen-like curve (Fig.3) turning the microwave on won't spark plasma in the chamber. However, if we put an identical sample in a 125 ml flask at the same 2000 mT pressure and turn the microwave on, the substrate can be treated with both microwave and plasma, thus giving us a way to differentiate the effects of microwave plasma quantitatively.

and is not an environmentally friendly process. Some recent works have shown alternative optical and thermal processes for obtaining reduced graphene oxide using laser[61,62]. Here we show that plasma generated by our simple system is capable of providing the local energy

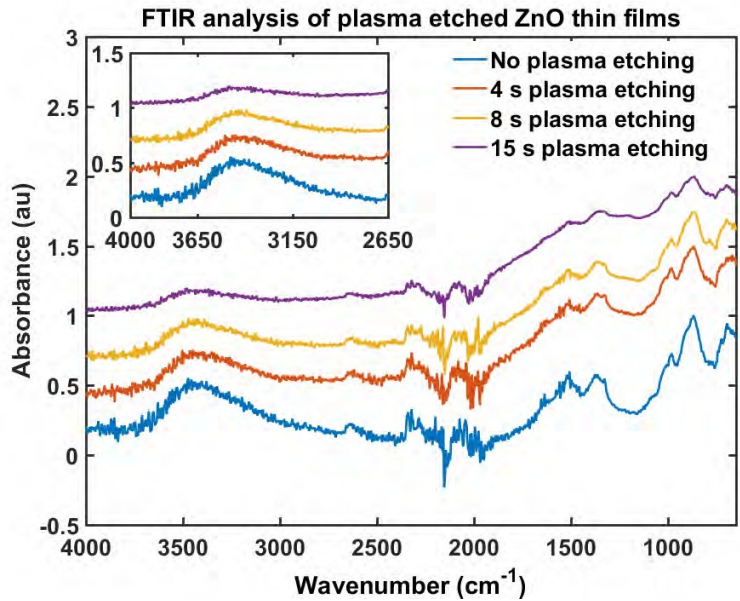

FIG. 7. The figure shows the result of plasma etching a ZnO nanowire thin film. The film was fabricated as for all devices (deposition from ethanol suspension) followed by sintering. The etching was conducted for 4, 8 and 15 seconds. The inset showing the band at 3400 cm$^{-1}$ due to the -OH bond in alcohols. The intensity of this signal clearly diminishes as the film is etched indicating a loss of electron-withdrawing groups from the surface of the material. The large signal at around 2100 cm$^{-1}$ is from $CO_2$ and the band at around 900 cm$^{-1}$ is due to a small amount of glass particles that were scraped off of the device substrate. The graphs shown above are the FTIR analysis of ZnO nanowire thin films, however other ZnO thin films made of ZnO particles or nanoparticles also show similar behaviour under plasma etching.

needed to reduce dielectric graphene oxide films to conductive graphene (Fig.Raman).

First, graphene oxide (Graphenea Inc.) is drop casted onto polyethylene terephthalate (PET) film. After drying for 24 hours at 50 °C, the GO-coated foil is cut into squares. The sheet resistance of each square is measured using a Keithley Source Meter with a four-point Kelvin probe. The resistance prior to plasma etching show a large value on the order of 10 MΩcm$^{-2}$.

One graphene oxide square is taped on either end to a glass substrate and then placed in the plasma etching flask. It is then evacuated for 1 minute (to a pressure of 300 mTorr). The flask is placed in the microwave for 4 seconds. The other two squares are also treated for 4 seconds, but neither affixed to substrates and one treated for 4 and 1 second intervals instead of continuous exposure. The third is exposed to plasma for 4 seconds continuously. The three types are referred to as supported, bursts, and free-standing, respectively as shown in Fig.9.

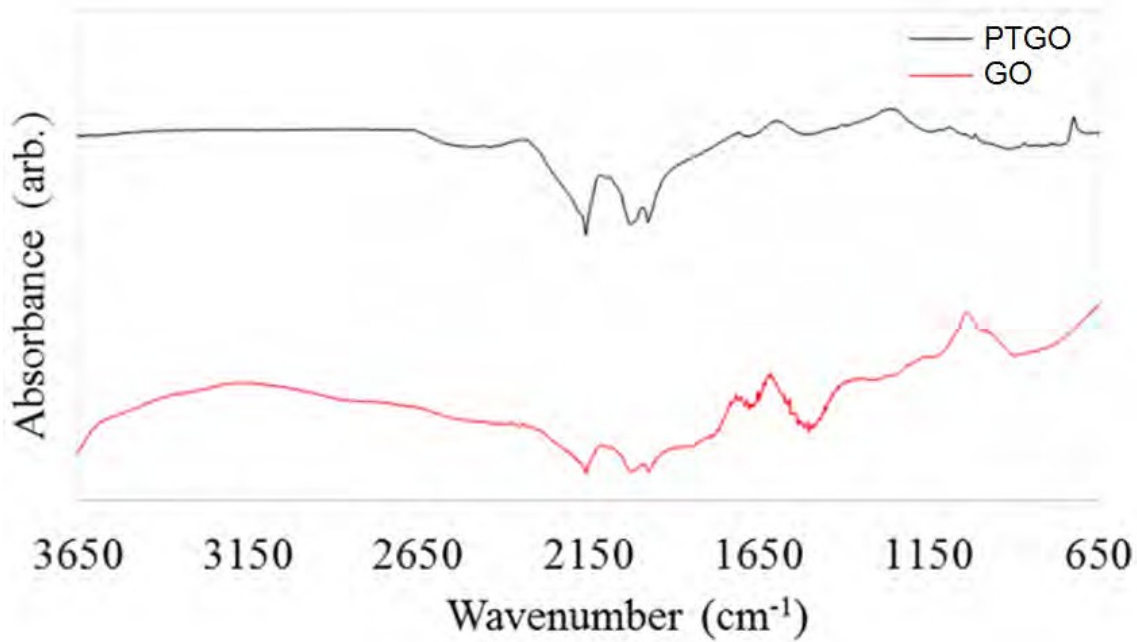

FIG. 8. Raman spectra and absorption studies for graphene oxide (GO) and plasma treated graphene oxide (PTGO) films.

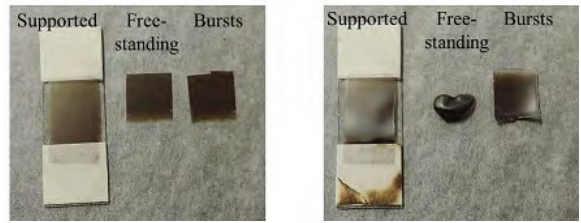

FIG. 9. Graphene oxide films before (L) and after (R) treatment in 4 seconds of air plasma. Clearly the free-standing sample is not very useful for applications despite is large change in resistance. A similar change in resistance was attained by treating the bursts-sample with four separate 1-second pulses of plasma, which prevented the plastic substrate from warping significantly.

The resistance following plasma etching is measured and showed a reduction of resistance of about two orders of magnitude. The lowest final resistance is obtained for the free-standing sample, but the heat from the plasma caused the plastic foil to warp, making this

| Substrate | Initial resistance $R_i$ (M$\Omega$cm$^{-2}$) | Final resistance $R_f$ (M$\Omega$cm$^{-2}$) | $\Delta R$ (%) |
|---|---|---|---|
| Supported | 5.4 | 0.2 | 96.3 |
| Free-standing | 10.5 | 0.1 | 99.0 |
| Bursts | 7.9 | 0.1 | 98.7 |

process less useful for flexible electronic applications.

### B. Bonding of Microfluidic Channels to Substrates

Many materials like polypropylene (PP), polyether ether ketone (PEEK), PDMS or polyoxymethylene (POM) are extremely hard to bond with other materials. Requirement of high bonding strength, durable and irreversible bonding of metal, glass and plastics present special challenges for the manufacturing industry. Plasma treatment of a surface with other applied cleaning procedures produce better adhesion capability and bonding strength on the surfaces to be joined in comparison to untreated substrates.

Microfluidics is an emerging technology with fluid flow in micrometer or less size channels and has tremendous applicability[63]. Applications for microfluidics include mobile chemical analysis[64], medical diagnostics[65], drug delivery[66], soft robotics[67], RNA encapsulation[68], and nanomaterials synthesis[69] to name a few. In order to fabricate robust microfluidic devices, it is necessary to have a strong bond between the microfluidic polymer mold (typically PDMS) and the substrate. A strong bond prevents the device from leaking, becoming damaged, or the development of passages by which the confined solutions could escape the network of channels. Plasma has been used in this application extensively, as it converts the exposed surface of the PDMS into dangling silane groups, allowing a very strong bond to glass to be formed. We demonstrate that the plasma generated in our simple microwave system can be used for this purpose.

The following steps are taken to create PDMS microfluidic channels bonded on glass: (1) A microchannel mold is created (typically using photolithography or other methods). (2) A mixture of liquid PDMS and cross-linking agent is then mixed in the 10:1 ratio and poured into the mold in a petri dish. The petri dish is kept in an incubator at 65 °C for a couple of hours. (3) The hardened PDMS is taken off the mold. A replica of the microchannels is obtained on the PDMS block. (4) To complete the microfluidic chip and to allow the injection of fluids for future experiments, the inlets and outlets of the microfluidic device are punched with a biopsy puncher whose diameter is slightly less than the size of the tubes to be connected. This will ensure tight fitting of the inlet/outlet tubes to the channels. (5) Finally, the side of the PDMS with open microchannels and the glass slide are treated with plasma to obtain closed channels with one flat inner wall of glass and all other walls of PDMS. (6) The plasma treatment irreversibly bonds PDMS with glass and makes the microfluidic chip.

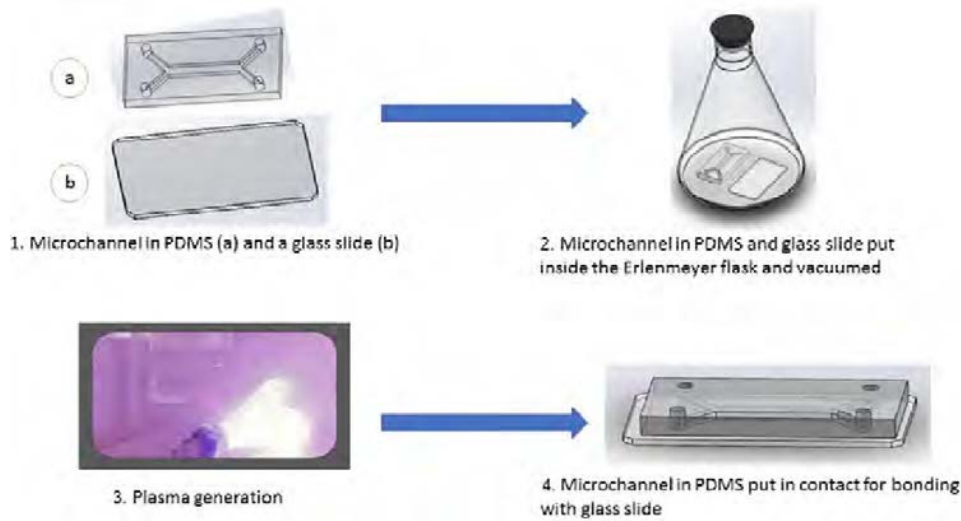

FIG. 10. A schematic to bond PDMS mold to glass to make microchannels.

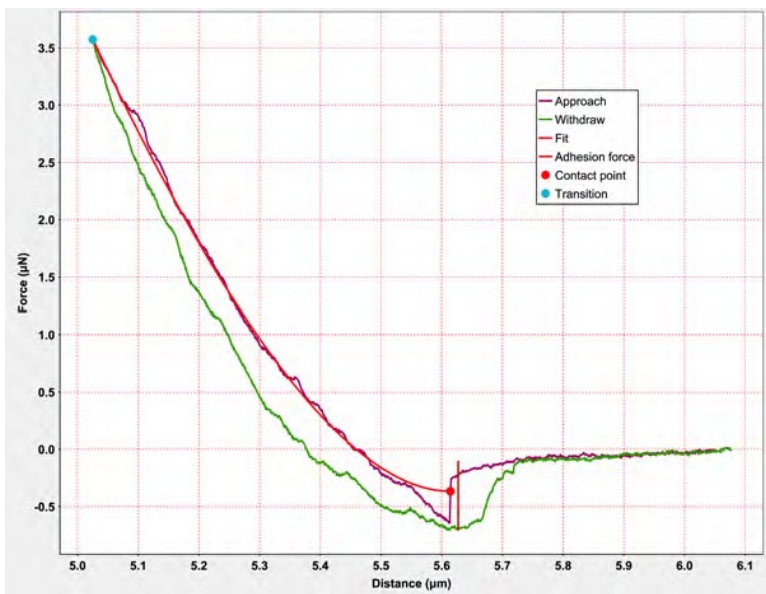

FIG. 11. Adhesive force of the untreated PDMS surface is measured using AFM force spectroscopy method.

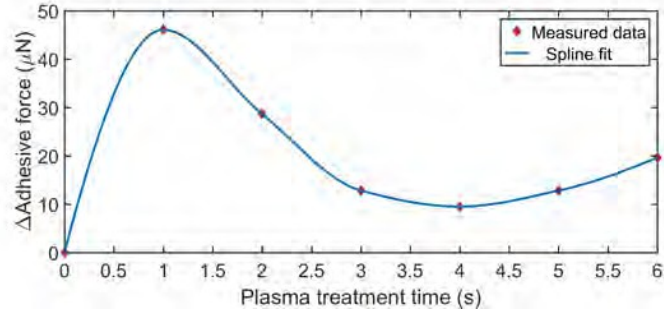

FIG. 12. Adhesive force of the PDMS surface is measured using AFM force spectroscopy method before and after various degrees of plasma treatment.

A freshly made PDMS replica and a glass microscope slide in two different vacuum flasks were used in this experiment. The flasks were evacuated to 500 mTorr (0.5 Torr) and three seconds of plasma exposure was used on both of them. After plasma treatment, the glass slide was removed and placed on a flat surface. The PDMS sample was then removed and the treated surface is placed in the desired location on the glass slide; the two components were then pressed together gently and allowed to sit for ten minutes. A schematic of the whole process is shown in Fig.10.

We have also measured the adhesive force of a PDMS surface before and after treatment using force-distance microscopy mode of an atomic force microscope (Nanosurf C3000 Flex-AFM) and the representative graph is shown in Fig.11. The relative change in adhesive force due to surface plasma treatment is shown in Fig.12.

## IV. CONCLUSION

In this work we have demonstrated that a simple plasma etching device can be constructed from a household microwave oven and a vacuum flask. We have showed that this apparatus can be used in surface treatment of a substrate for several cutting-edge research applications. Varying degree of plasma exposure to a PDMS substrate leads to change in surface energy and hence a change in contact angle of a water droplet on the substrate was observed before and after the surface treatment. Upon exposure to plasma a change in opto-electrical properties of a metal oxide semiconductor such as ZnO was demonstrated using FTIR spectra analysis. Significant change in electrical resistance of graphene oxide thin films was also observed as an effect of plasma exposure. Moreover, it was also shown that irreversible bonding between certain elastomers like PDMS and glass can be achieved to make microfluidic channels with this simple technique. Needless to say that the domain of these applications is enormous and our hope is that this paper, through this frugal alternative to conventional and expensive plasma treatment, would enable more researchers to delve into these cutting edge research areas leading to more innovation and discovery.

## V. ACKNOWLEDGEMENT

This work was partially supported by the National Science Foundation (HBCU-UP Award # 1719425), the Department of Education (MSEIP Award # P120A70068) with MSEIP CCEM Supplemental award and Maryland Technology Enterprise Institute through MIPS grant. KD would like to thank Dr. Jim Marty of Minnesota Nano-Science Center and NanoLink for providing support and materials for photolithography process in microchannel fabrication and Dr. Aaron Persad of MIT for many helpful discussions and suggestions. We also thank MIT Technology Review for featuring this work[70].

# SUPPLIMENTARY MATERIALS

We are also using the plasma generation using a kitchen microwave experiment as a classroom demonstration in an effort to engage students in the discussion on some of the important concepts in electromagnetism in undergraduate physics. We have created a pool of multiple choice concept cartoon questions to initiate discussions during classroom lectures on important concepts such as the nature of electromagnetic radiation, relationship between force/kinetic energy of a charge and the applied electrical field, dipole moment vector and the torque applied by an oscillating electric field, collission cross section and cascading effect to generate plasma etc. Impact of similar cartoon questions were measured and reported[71]. High resolution pictures of these slides, i.e., Fig.13-Fig.30, can be obtained from the corresponding author of this paper.

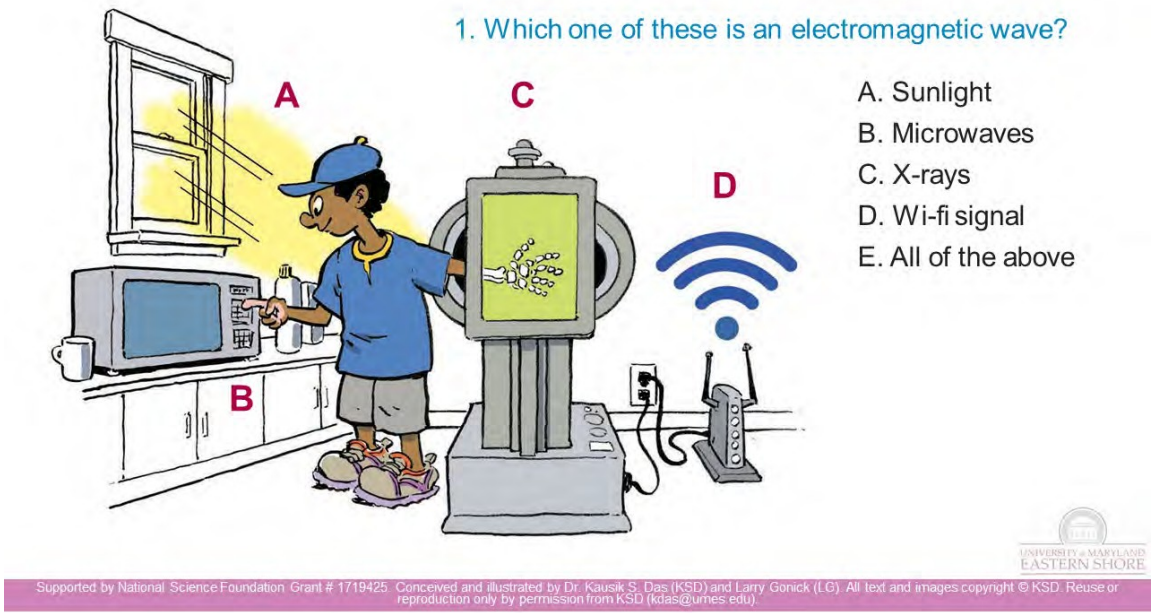

FIG. 13. This slide elaborates that we are surrounded by the electromagnetic waves.

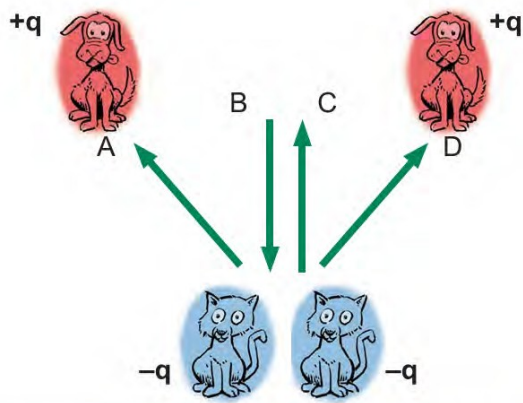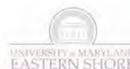

FIG. 14. This slide can be used to elaborate the concept of dipole moment vector.

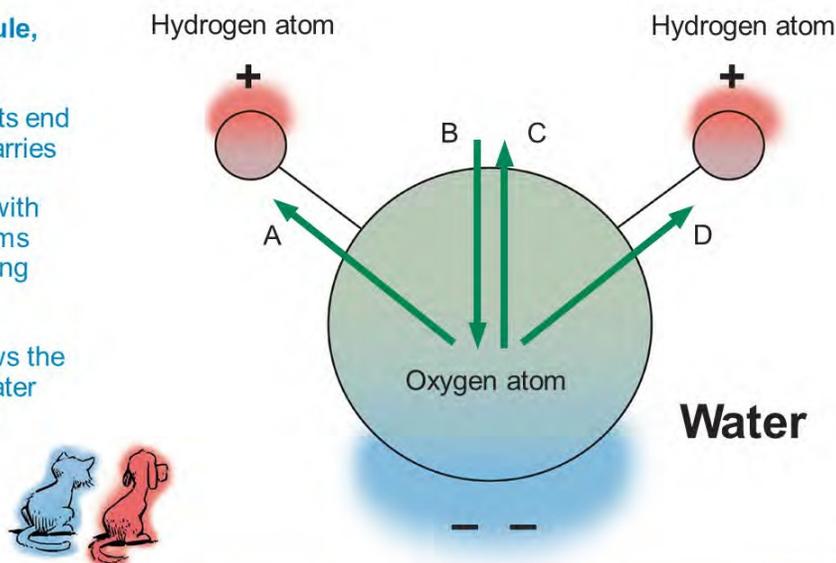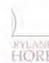

FIG. 15. Dipole moment of water molecules.

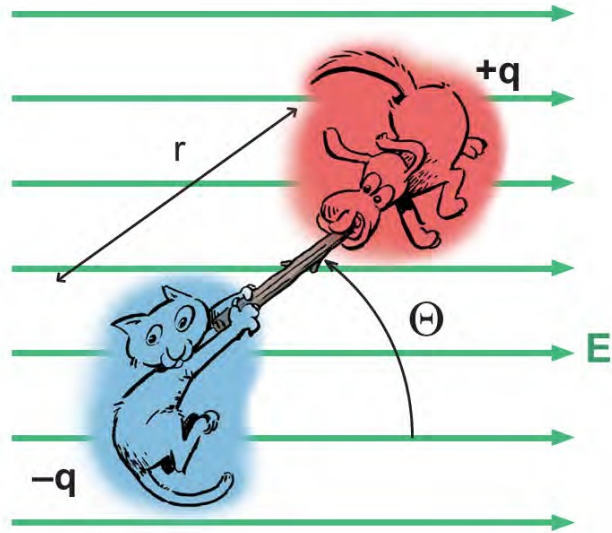

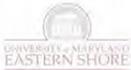

FIG. 16. This slide leads to the discussion of torque applied by an electrical field on a dipole.

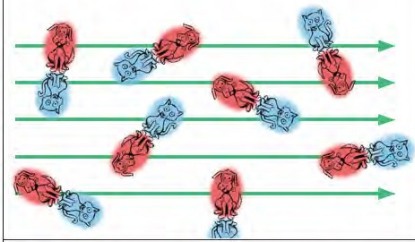
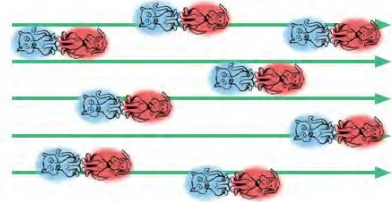
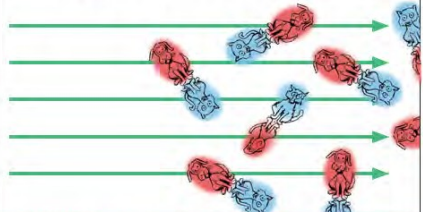
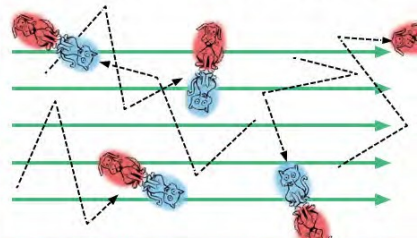

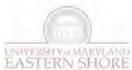

FIG. 17. Allignment of a group of dipoles along the direction of a steady electrical field.

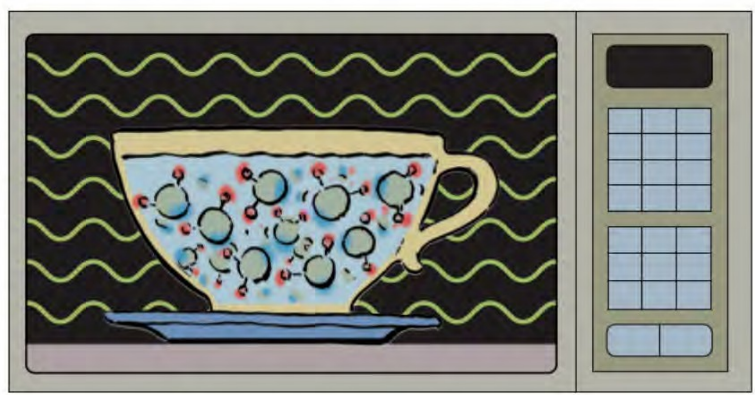

FIG. 18. This slide elaborates oscillation of a dipole in an oscillating electric field.

FIG. 19. This slide illustrates that in a kitchen microwave oven water molecules oscillate in an oscillating electric field. However, due to hydrogen bonding between the molecules and viscosity of water the oscillating molecules rapidly dissipate energy, in turn heating up the substrate.

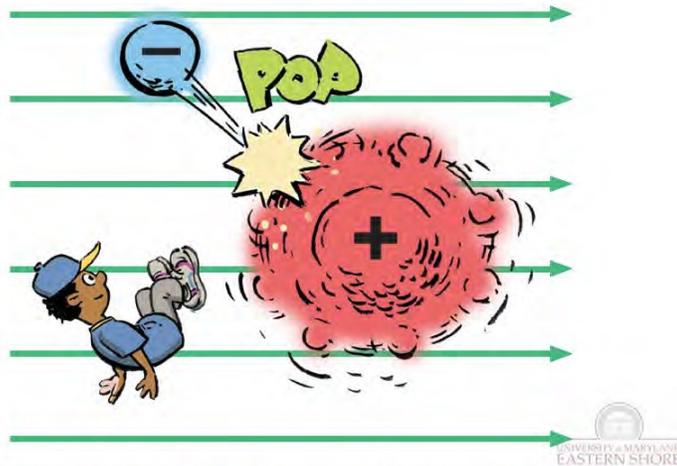

FIG. 20. One possible explanation to generate primary electrons. However, as discussed in Section II.B, more work is needed to confirm this scenario.

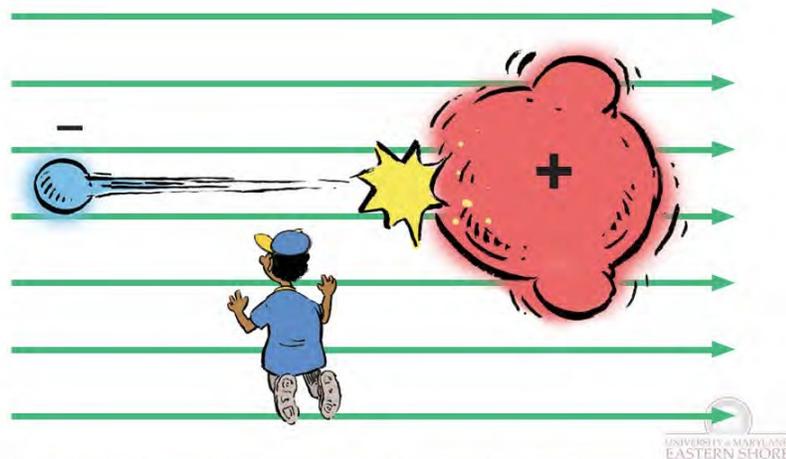

FIG. 21. Effect of instantaneous electric field on positive and negatively charged particles.

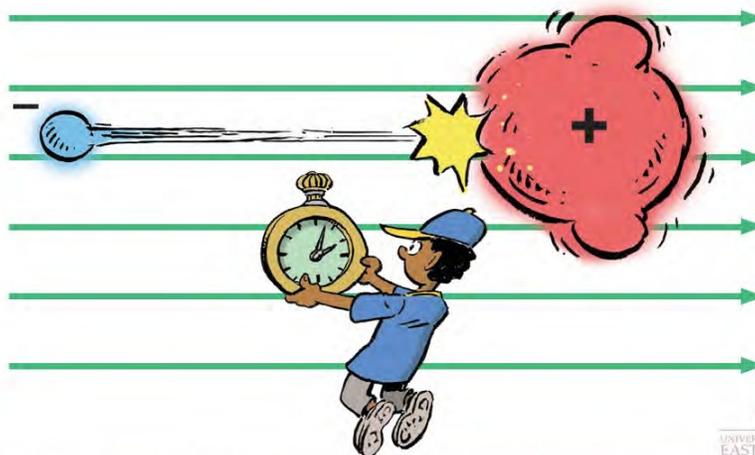

FIG. 22. This slide can be used to initiate discussions on the effect of mass of the charged particles under the influence of an electric field on their kinetic energies after equal time interval.

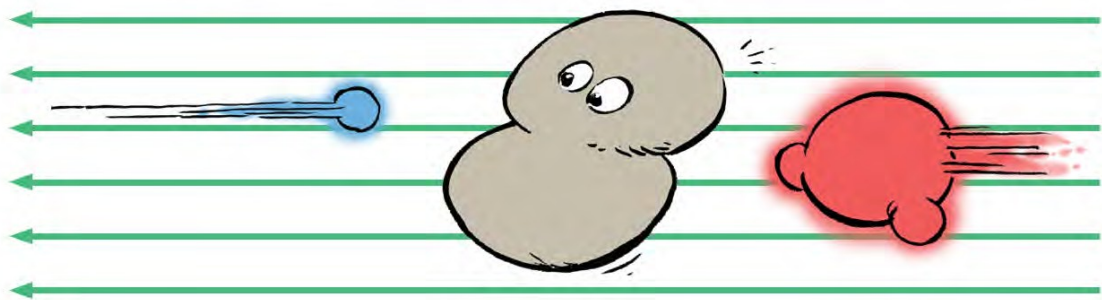

FIG. 23. This is a follow up question on the effect of mass of the charged particles under the influence of an electric field on their kinetic energies.

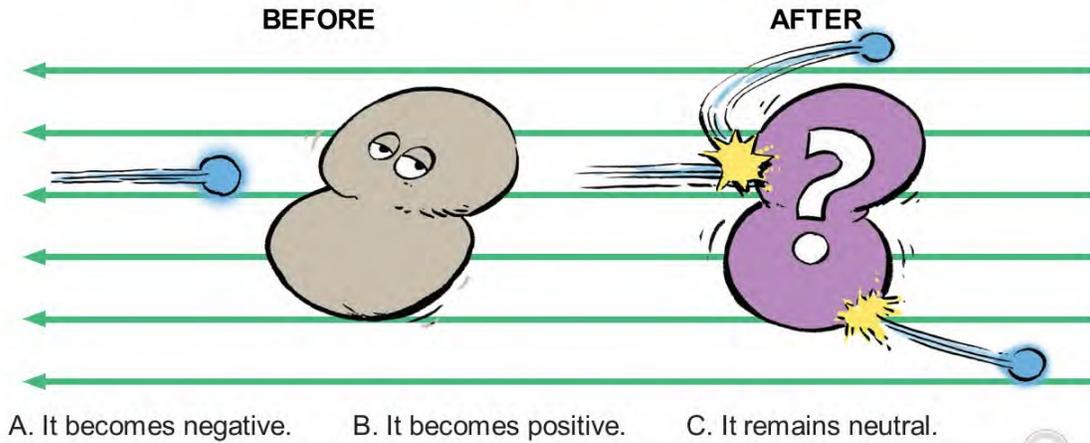

FIG. 24. This slide demonstrates the mechanism of secondary electron generation and ionization of an atom/molecule by the primary electron.

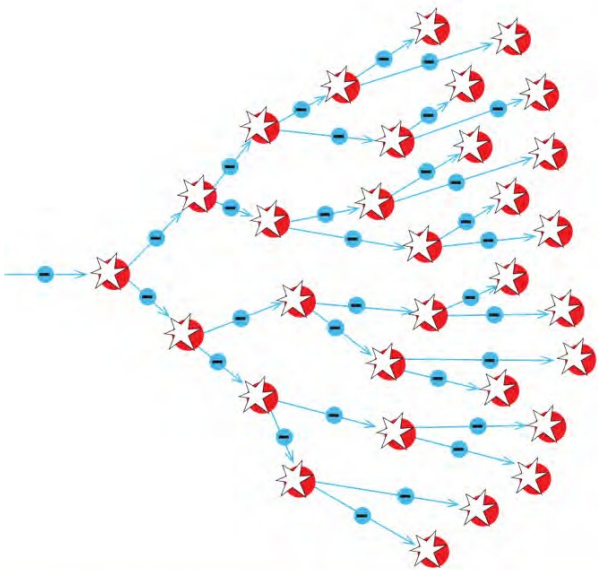

FIG. 25. This slide demonstrates the chain reaction and the path towards plasma generation by Townsend type avalanch effect.

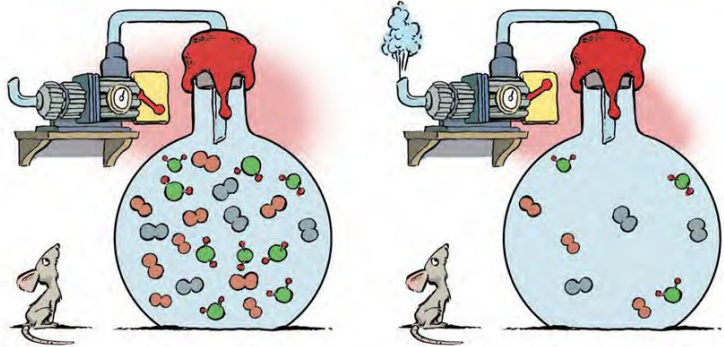
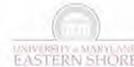

FIG. 26. This slide demonstrates the effect of gas pressure on the mean free path.

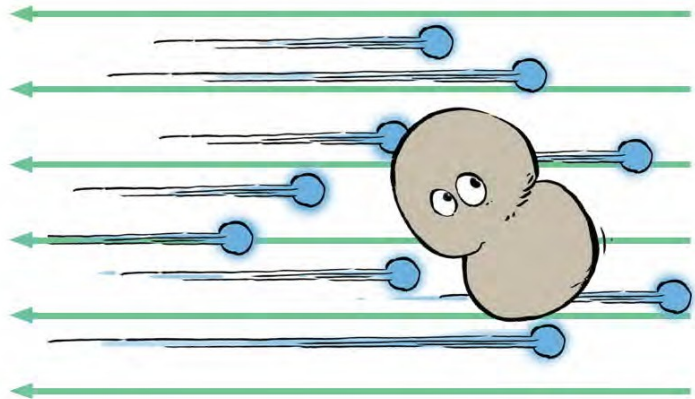
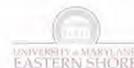

FIG. 27. This slide demonstrates that longer mean free path means more time for the electrons to gain more kinetic energy before the next collision.

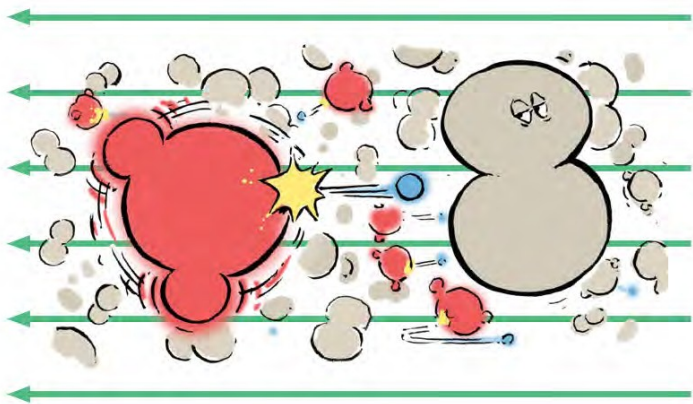
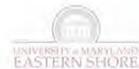

FIG. 28. This slide demonstrates that shorter mean free path means less kinetic energy for the electrons before the next collision.

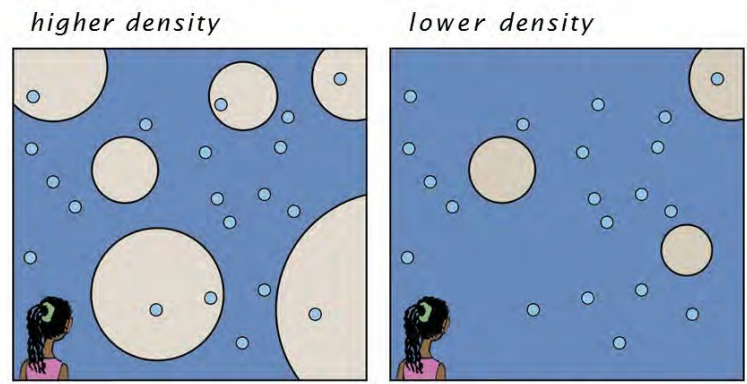
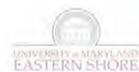

FIG. 29. This slide demonstrates that effect of pressure on the collision cross section of electrons with other constituent gas molecules.

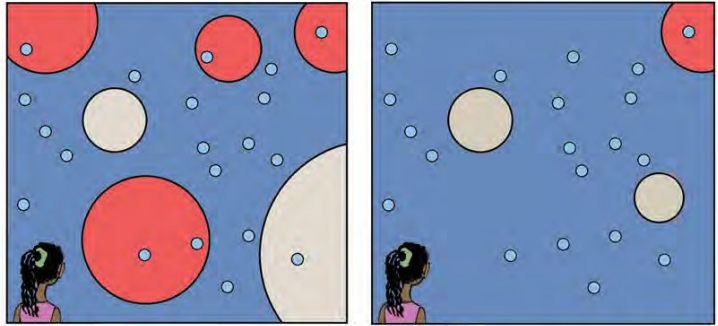

FIG. 30. This slide initiates a discussion on how decreasing gas pressure may decrease collision cross section and in turn reduce the probablity of generating higher order electrons in turn adversely affecting the possibility of plasma generation.

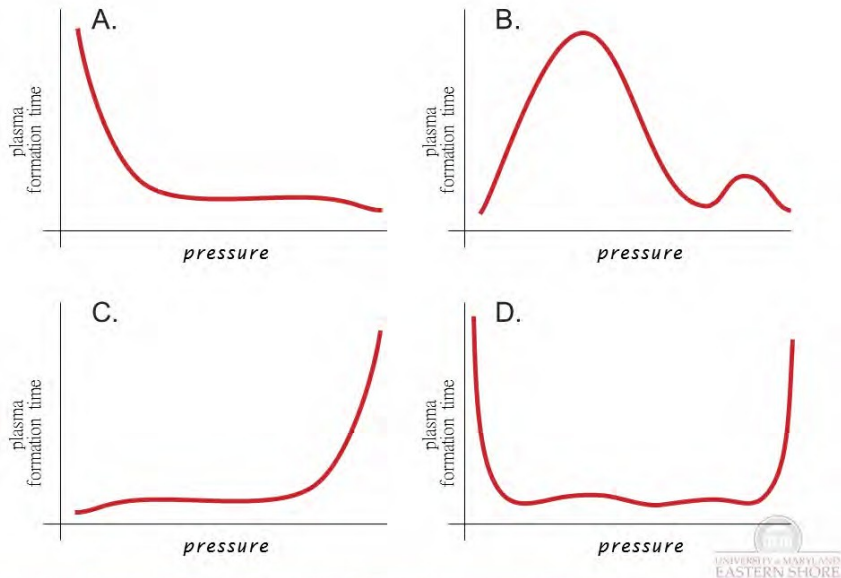

FIG. 31. This slide initiates a discussion on the assymptotic nature of plasma initiation with respect to the pressure of the gas. We have seen that at constant temperature, very low gas pressure reduces collision cross section, while high gas pressure reduces electrons average kinetic energy in between two successive collisions. Thus there should be a sweet spot between these two competing parameters where the condition for plasma generation is just right.